\RequirePackage{lineno}
\documentclass[article,prl,
twocolumn,
amsmath,amssymb,
amsmath,amssymb, aps,
]{revtex4}
 
\usepackage{epsfig}
\usepackage{amsfonts}
\usepackage{amsmath}
\usepackage{amssymb}
\usepackage{bm} 
\usepackage[framemethod=PStricks]{mdframed}
\usepackage{lipsum}
\usepackage{float}
\usepackage{color}
\usepackage{pdflscape}
\usepackage{graphicx}

\newcommand{\be}{\begin{equation}}
\newcommand{\ee}{\end{equation}} 
\newcommand{\bea}{\begin{eqnarray}} 
\newcommand{\eea}{\end{eqnarray}}

\usepackage[utf8]{inputenc}
\usepackage{bbm} 				
\usepackage{epstopdf}				
\usepackage{verbatim}    			
\usepackage{sidecap}                
\usepackage{indentfirst}
\usepackage[colorinlistoftodos]{todonotes}

\begin{document}

\title{Magnetic Dissipation of Near-Wall Turbulent Coherent Structures \textcolor{black}{in Magnetohydrodynamic Pipe Flows}}

\author{L. Moriconi}
\affiliation{Instituto de F\'\i sica, Universidade Federal do Rio de Janeiro, \\
C.P. 68528, CEP: 21945-970, Rio de Janeiro, RJ, Brazil}


\begin{abstract}
Relaminarization of wall-bounded turbulent flows by means of external static 
magnetic fields is a long-known phenomenon in the physics of electrically conducting
fluids at low magnetic Reynolds numbers. Despite the large 
literature on the subject, it is not yet completely clear what combination of the 
Hartmann ($M$) and the Reynolds number has to be used to predict the laminar-turbulent 
transition in channel or pipe flows fed by upstream turbulent flows free of magnetic 
perturbations. Relying upon standard phenomenological approaches related to mixing 
length and structural concepts, we put forward that $M/R_\tau$, where $R_\tau$ is the 
friction Reynolds number, is the appropriate controlling parameter for relaminarization, 
a proposal which finds good support from available experimental data.

\end{abstract}


\maketitle



The laminar-turbulent transition of electrolyte or liquid metal flows at low magnetic Reynolds numbers 
($R_m \ll 1$), where external magnetic fields are negligibly affected by {\textcolor{black}{induced currents}}
\cite{moreau, knaepen, davidson, verma}, is a challenging scientific problem of great technological interest.
As it was discovered long ago by Hartmann and Lazarus \cite{hart_laza}, turbulent flows of electrically conducting
fluids can be relaminarized when subject to large enough magnetic fields (for an interesting set of numerical simulations
see \cite{num1,num2,num3,num4,num5}). 
{\textcolor{black}{Relaminarization from the action of electromagnetic fields holds, as a matter of fact, 
on a broader context, not necessarily within the framework of low $R_m$ flows.
The general idea is found in studies of flow control \cite{tsinober1, spacecraft}}}, 
in semiconductor crystal growth \cite{scrystal, chen_etal}, 
in the design of tritium breeding blankets for fusion reactors \cite{tritiumb, ihli_etal}, and in steel 
casting \cite{davidson2,steelc}. In all of these domains, the essential issue is to understand whether laminar 
or turbulent flow regimes will take place under the presence of magnetic fields for a variety of boundary conditions.

Flow control by magnetic fields is a promising strategy for the thermal protection of spacecrafts
during atmospheric reentry ({\textcolor{black}{a high $R_m$ case}), while laminar flows of melted steel are welcome in casting processes in order to avoid particulate entrainment at the steel-air interface, and also in semiconductor growth technology, for homogeneity enhancement in the production of silicon ingots. In fusion research, in contrast, one is interested to prevent turbulence attenuation of the coolant flow (one of the roles of tritium blankets) 
associated to the strong magnetic fields produced by the fusion plasma.

Dynamic similarity for incompressible electrically conducting flows is parameterized by two dimensionless 
quantities, the Hartmann ($M$) and Reynolds ($Re$) numbers \cite{moreau, knaepen, davidson, verma}. 
{\textcolor{black}{More precisely,
$M^2$ and $Re$ are estimates, respectively, of the ratio of magnetic and inertial forces to viscous forces. }}
{\textcolor{black}{It turns out that in the asymptotic limit of high transverse magnetic fields, pipe flows become laminar with drag coefficients proportional to $M/Re$. A word of caution is in order here, insofar the laminar-turbulent transition is not a universal phenomenon at high $M/Re$, as it is known from studies of flows subject to other boundary conditions \cite{eckert,potherat-klein2017,potherat-kornet}.} 


The original discussions presented in the seminal papers \cite{hart_laza, murga, lykoudis, brou_lyko} 
have proposed that $M/Re$ should work as a controlling parameter for the laminar-turbulent transition 
--  an educated guess that still percolates in much of the recent literature. However, broadly 
acknowledged and careful pipe flow experiments performed by Gardner and Lykoudis \cite{gardner_lyko} almost 
fifty years ago, established that critical values of $M/Re$ actually depend on the Reynolds 
numbers at the laminar-turbulent transition point, a fact emphasized by Tsinober \cite{tsi} 
and further discussed by Branover \cite{bran}, who proposed that a more accurate transition 
criterion would be given by $M/Re^\alpha$, where the exponent $\alpha$ is slightly smaller 
than unity. Narasimha pointed out, subsequently, that relaminarization could be related to 
flow regimes where magnetic forces dominate Reynolds stress gradients \cite{nara}. 

{\textcolor{black}{Our aim in this paper is to model the Gardner-Lykoudis 
measurements of the laminar-turbulent transition, having in mind contemporary
ideas about the role of coherent structures in wall-bounded flows.
They have been introduced as fundamental modeling elements in the structural 
approach to hydrodynamic turbulence \cite{townsend1, theo,townsend, perry_chong, 
perry_marusic, marusic_perry,review1,review2,review3} and have, similarly, 
been the focus of great attention in magnetohydrodynamics, as in the investigation of the dynamo effect \cite{tobias_etal}, 
particulate deposition in duct flows \cite{liu_etal},
magnetic reconnection \cite{loureiro_etal}, and solar wind heating \cite{camporeale_etal}, to name just 
a few examples out of a myriad of studies.}}



The dynamic evolution of a neutral fluid of mass density $\rho$, dynamic 
viscosity $\mu$, and conductivity $\sigma$, which is subject to a static uniform magnetic 
field $\vec B$, is governed, at low $R_m$ (when fluctuations of the magnetic field are quickly damped), by the electromagnetically forced Navier-Stokes equations \cite{moreau, knaepen, davidson, verma}, 

\be
\partial_t \vec v + \vec v \cdot \vec \nabla \vec  v  =  - \vec \nabla P + \nu \nabla^2 \vec v + 
 \frac{\sigma}{\rho} \left ( - \vec \nabla \phi + \vec v \times \vec B \right ) \times \vec B \ , \  \label{mhd1}
\ee
where the usual incompressibility constraint, $\vec \nabla \cdot \vec v = 0$ is imposed,
$\nu = \mu / \rho$ is the kinematic viscosity, and the electric potential $\phi$ 
satisfies the Poisson's equation
\be
\nabla^2 \phi = \vec \nabla \cdot (\vec v \times \vec B) \ . \ \label{phi}
\ee
Considering a statistically stationary flow in a pipe of diameter $D$, with bulk velocity $U$, the dynamic 
equations can be rewritten in dimensionless form with the help of the following substitutions,
\be
\vec v \rightarrow U \vec v \  , \ P \rightarrow  \nu \frac{U}{L} P \ , \ 
\phi \rightarrow UD \phi \ , \ \vec r \rightarrow D \vec r \ , \ t \rightarrow \frac{L^2}{\nu} t \ . \
\ee
Eq. (\ref{mhd1}) becomes, then,
\be
\partial_t \vec v + Re (\vec v \cdot \vec \nabla \vec  v)  =  - \vec \nabla P + \nabla^2 \vec v + 
 M^2 \left ( - \vec \nabla \phi + \vec v \times \hat B \right ) \times \hat B \ , \  \label{mhd2}
\ee
where
\be
Re \equiv \frac{UD}{\nu} \label{reynolds}
\ee
is the Reynolds number for the flow, $\hat B \equiv \vec B / B$ is the versor parallel to the magnetic field, 
and
\be
M \equiv B D \sqrt{\frac{\sigma}{\mu}}
\ee
is the Hartmann's number for the magnetohydrodynamic system.
In typical low $R_m$ experiments which probe the laminar-turbulent transition, a pipe or channel 
turbulent flow enters a region subject to a uniform magnetic field.

{\textcolor{black}{We may draw a couple of important and general results for low $R_m$ flows, by just relying on energy budget and vorticity equations. Assume that the pipe flow takes place in a long pipe of length $L$ with periodic boundary conditions for the inlet and outlet velocity fields, the electric potential, and the pressure fluctuation field, in such a way that the inlet and outlet flow rates are guaranteed to be the same and no mean streamwise electric field arises along the pipe. The flow is driven by a constant, non-fluctuating, background pressure gradient, $\Delta P / L$. Integrating, now, the scalar product of Eq. (\ref{mhd1}) with the velocity field over all the pipe domain and formally writing the solution of the Poisson's Eq. (\ref{phi}) in terms of the inverse Laplacian operator as (boundary conditions for the electric potential are tacitly assumed),
\be
\phi(\vec r,t) = \nabla^{-2} [ \vec \nabla \cdot (\vec v \times \vec B ) ] \ , \
\ee
we get the energy budget equation
\be
\dot E =   \Phi \Delta P + \dot E_B^{(+)} - \dot E_B^{(-)} - \dot E_\mu \ , \  \label{dotE}
\ee
where $\Phi$ is the pipe's flow rate,
\be
\dot E_\mu =  \mu \int d^3 \vec r (\partial_i v_j)^2
\ee
is the viscous dissipation rate, while
\be
\dot E_B^{(+)} = - \sigma \int d^3 \vec r (\vec \omega \cdot \vec B) \nabla^{-2} (\vec \omega \cdot \vec B)  \label{E+}
\ee
and
\be
\dot E_B^{(-)} =  \sigma \int d^3 \vec r [ \vec v^2 \vec B^2 - (\vec v \cdot \vec B)^2] \label{E-}
\ee
provide energy input and Ohmic dissipation rates, respectively, associated to the Lorentz forces. Above, 
$\vec \omega = \vec \nabla \times \vec v$ is the vorticity field.}}

{\textcolor{black}{The energy dissipation rates $\dot E_\mu$ and $\dot E_B^{(-)}$ are always positive (energy sink contributions), whereas $\dot E_B^{(+)}$ is in general positive and vanishes only for vorticity fields which are perpendicular to the background magnetic field.}}

{\textcolor{black}{
Eqs. (\ref{E+}) and (\ref{E-}) can be combined to give the overall (non-viscous, non-positive definite) energy production rate, 
$ \dot E_B  \equiv \Phi \Delta P + \dot E_B^+ - \dot E_B^- $, with
\be
\dot E_B^+ - \dot E_B^-   = \sigma \int d^3 \vec r [ (\vec v \cdot \vec B)^2 + (\vec \omega_\perp \cdot \nabla^{-2} 
\vec \omega_\perp ) \vec B^2] \ , \ \label{E_B}
\ee
where $\vec \omega_\perp$ is the  projected vorticity field on the plane normal to $\vec B$. Relation (\ref{E_B}) indicates that vortices are dissipated on time scales that depend on their core sizes, as well on their orientations, as it is in fact expected from the discussions presented in Refs. \cite{sommeria-moreau1982,davidson1997,potherat-klein2014}. It is also clear that vortices which are aligned with the magnetic field ($\vec \omega_\perp = 0$) contribute to the energy bugdet equation (\ref{dotE}) with positive energy production rates that could even affect the global energy balance of the flow. One may wonder, then, why these vortices are suppressed in usual laminarized pipe or channel 
low $R_m$ flows. To address this point, it is convenient to write down the dynamic equation for the vorticity field. Taking the curl of {\hbox{Eq. (\ref{mhd1})}}, we get
\be
\frac{D \vec \omega}{Dt} =  \vec \omega \cdot \vec \nabla \vec  v + 
 \nu \nabla^2 \vec \omega + 
\frac{\sigma}{\rho} \left [\vec \nabla (\vec v \cdot \vec B ) \times \vec B - \vec \omega \vec B^2  \right ] 
\ , \ \label{dot_omega}
\ee
where $D/Dt = \partial / \partial t + \vec v \cdot \vec \nabla$ is the material derivative operator. The first term between brackets in the RHS of the above equation implies that 
fluctuations of {\hbox{$\vec v  \cdot \vec B$}} lead, eventually, to 
vortex tilting toward planes that are perpendicular to $\vec B$. 
As a general consequence, the contribution of 
vortical structures to the injected power $\dot E_B^{(+)}$ gets damped, up to the point where they are
completely swept out from the flow. The mechanism just alluded for vortex suppression is related, in an alternative geometric context and at asymptotically high magnetic fields, to the dissipation of vortices that cross the so-called characteristic surfaces \cite{moreau, alboussiere_etal1996, moreau_etal2010}, two-dimensional manifolds that contain streamlines (or currents) and magnetic field lines.}

{\textcolor{black}{The decay of magnetically aligned vortices, however, may happen to be
not so fast along the flow, so that they can be occasionally detected at the outlet of relatively long channels 
in conditions of expected laminar regimes \cite{suko_etal,zika_etal}. This phenomenon can be of particular relevance in channel flows, in contrast to the case of pipe flows, since, in the latter, characteristic surfaces are more often crossed, due to purely geometric reasons, by near-wall vortices that are lifted by ejection events toward the bulk flow \cite{adrian2007}.}} 

{\textcolor{black}{It is interesting to call attention, furthermore, to the fact that if the flow is subject to 
additional external forces or obstacles that produce vorticity along the magnetic 
field direction, the vortex tilting mechanism discussed above can be blocked. In this way, vortex structures parallel 
to the magnetic field can be sustained and unsuspected two-dimensionalized turbulent regimes can arise from the action 
of large magnetic fields \cite{eckert,potherat-klein2017}.}}

\begin{figure}[ht]
\hspace{0.0cm} \includegraphics[width=0.4\textwidth]{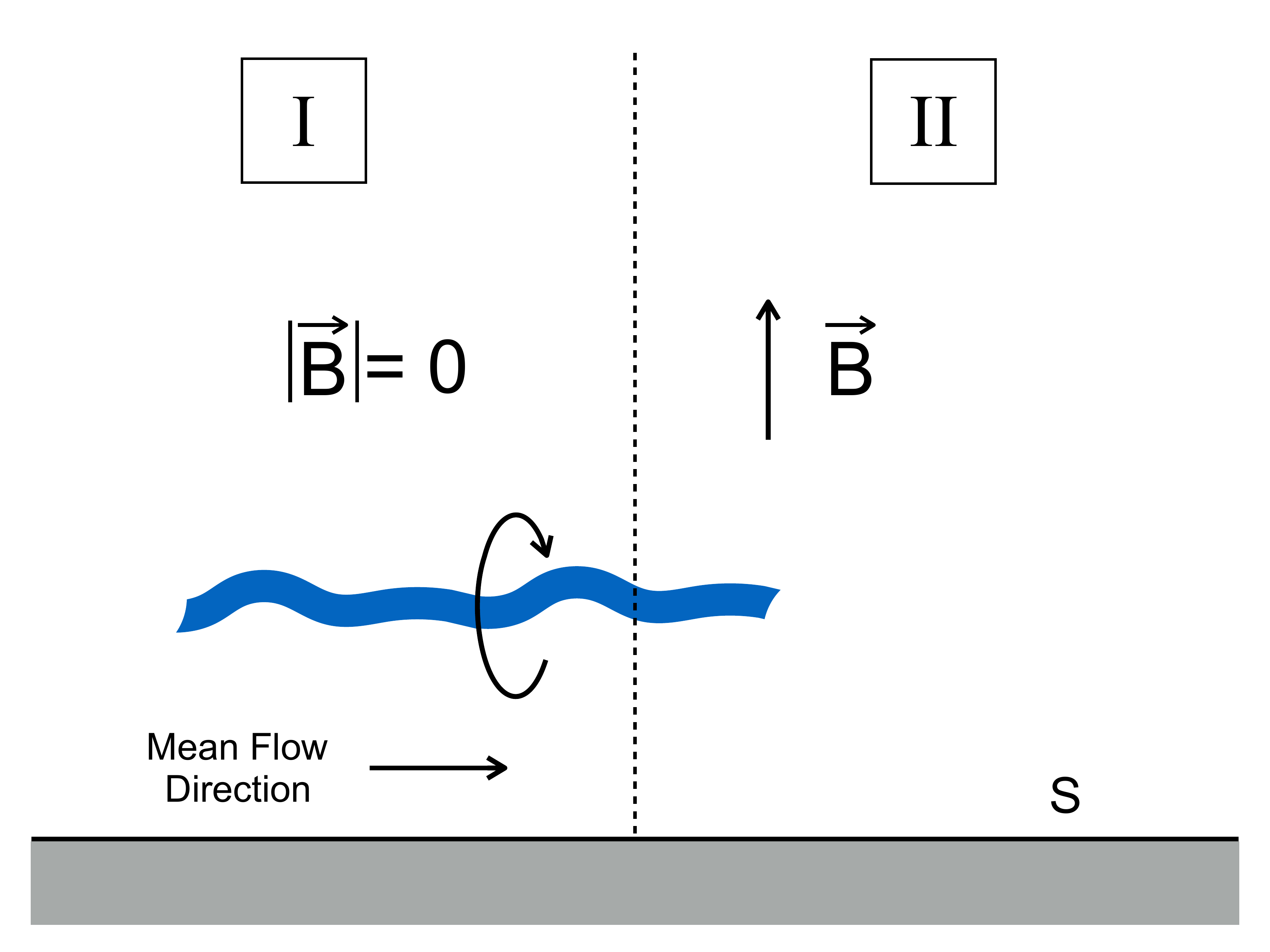}
\vspace{0.0cm}
\caption{Quasi-streamwise vortices, which are distant from the wall S by a few viscous lengths, 
move from region I, free of magnetic forces, to region II, where they are annihilated due to the 
presence of a static and uniform external 
magnetic field {\textcolor{black}{(to avoid misinterpretations, the transition between regions I and II is taken to be as smooth as the one produced by external currents, with $\vec \nabla \times \vec B = 0$ in the entire fluid domain).}} }
\label{vr}
\end{figure}

{\textcolor{black}{As it has been noticed in experiments and numerical simulations, as the Hartmann number grows, at fixed turbulent Reynolds number, 
the flow starts to become laminar in the core region of wall-confined flows \cite{gardner_lyko,num1,num2,num3,num4,num5,liu_etal}. Initially, since viscous forces are much smaller than inertial and magnetic forces in the bulk flow, {\textcolor{black}{the linear dimensions of the core laminar region, along directions which are transverse to the mean flow, are expected}} to depend on the interaction parameter $M^2/Re$ (also known as the Stuart number, the ratio of magnetic forces to inertial forces). The laminar core grows until viscous forces in its surroundings become dominant over the inertial ones. 
At the onset of laminarization, {\textcolor{black}{the incoming near-wall coherent structures, which are advected from the inlet flow (free of magnetic fields) and are}} observed to be mainly quasi-streamwise vortices \cite{jimenez,elsas_moriconi}, lose kinetic energy (and angular momentum, as well) due to the dissipative action of the magnetic field as they are carried by the 
mean stream. Therefore, the route to laminarization begins and ends with the suppression of outer and inner boundary layer mixing, respectively.}}

A sketch of the phenomenological picture addressed here is shown in {\hbox{Fig. 1}}. If the magnetic field is strong enough, 
quasi-streamwise vortices eventually disappear downstream and, as a consequence, vorticity fluctuations associated to vortical structures cannot be propagated anymore from the wall to the bulk of the flow, which, then, becomes fully laminar.

In order to get an asymptotic criterion for magnetic relaminarization, we assume, as
a reasonable simplification, {\textcolor{black}{that quasi-streamwise vortices are perfectly aligned with 
the wall, which is taken to be perpendicular to the magnetic field lines. In this case, from (\ref{E+})
and (\ref{E-}), it follows that $\dot E_B^{(+)} = 0$, and that the
power per unit volume provided by the magnetic force is}}
\be
P_B \equiv \sigma \left [ (\vec v \cdot \vec B)^2 - \vec v^2 \vec B^2 \right ] \leq 0 \ . \  \label{P_B}
\ee
The non-positiveness of $P_B$ tells us that quasi-streamwise vortices are in fact dissipated
by the transverse magnetic field. Referring back to {\hbox{Fig. 1}}, the vortical structures
that cross from region I to region II are furthermore supposed to have translation/core rotational 
velocities and linear dimensions that scale with the friction velocity $u_\tau$ and the viscous 
length $\ell$ \cite{jimenez,geng_etal}. The total power injected per vortex into region II can, therefore, be 
straightforwardly estimated as
\be
P_{in} \sim \rho u_\tau^3 \ell^2 \ . \
\ee
On the other hand, the power dissipated per vortex in region II, due to the presence of the 
magnetic field is analogously expected to be, from (\ref{P_B}),
\be
P_{out} \sim \sigma u_\tau^2 B^2 \ell^3 \ . \
\ee
A {\it{sufficient}} condition for relaminarization in pipe flow is naturally 
written as
\be
P_{out} \gg P_{in} \ , \
\ee
which leads to
\be
\frac{M}{R_\tau} \gg 1 \ , \  \label{crit}
\ee
where
\be
R_\tau \equiv \frac{u_\tau R}{\nu} = \frac{R}{\ell} \label{Rt} 
\ee
is the usual definition of the friction Reynolds number in pipe flows \cite{pope}.

Condition (\ref{crit}) can be derived, actually, from the evaluation of the ratio between convective and magnetic dissipative time scales in the very near-wall region. Still keeping an eye on Fig. 1, it is suggested that coherent structures produced by flow instabilities in the viscous layer of region I, near the interface between regions I and II, are transported to upper layers and dissipated by the magnetic field in region II, within the respective time scales
\be
T_C \sim \frac{\ell}{u_\tau} = \frac{\ell^2}{\nu} \label{TC}
\ee
and
\be
T_B \sim \frac{\rho}{\sigma B^2} \ . \ \label{TB}
\ee
A transition to the laminar flow regime is expected to take place whenever $T_C / T_B$ goes 
beyond a critical threshold. Note, in particular, that
\be
\frac{T_C}{T_B} \sim \left (\frac{M}{R_\tau} \right )^2 \ , \
\ee
which yields an alternative interpretation of (\ref{crit}). {\textcolor{black}{It is furthermore interesting to 
compare $T_C$ to the convective time scales of the inlet bulk turbulent flow, that is
\be
T_C' \sim \frac{R}{U} \ . \ \label{TCp}
\ee
We have, thus,
\be
\frac{T_C}{T_C'} \sim \frac{Re}{R^2_\tau} \sim \left ( \frac{U}{u_\tau} \right )^2 \frac{1}{Re} \ . \
\ee
Since $\left ( u_\tau / U \right )^2$ is proportional to the pipe friction factor \cite{pope}, 
it is not difficult to conclude, from the analysis of compiled friction (Moody) charts \cite{pope}, that the above quantity vanishes as $Re \rightarrow \infty$. This means that near-wall turbulent mixing takes places along time scales which are much shorter than the bulk ones. The estimates (\ref{TC}), (\ref{TB}), and (\ref{TCp}) suggest, thus, that the process of magnetic laminarization of turbulent 
channel or pipe flows is in fact expected to happen initially in their bulk (core) regions, as formerly pointed out.}}

{\textcolor{black}{A refinement of the above argument points to the relevance of the geometry and inlet boundary conditions on the nature of flow regimes at high Hartmann's numbers. Playing with Eqs. (\ref{E+}) and (\ref{E-}) for the case of straight axisymmetric vortices which make an angle $\theta$ with the magnetic field direction, one finds that 
\be
\dot E^{(+)}_B = a \cos^2 \theta 
\ee
and
\be
\dot E^{(-)}_B = a - b \sin^2 \theta \ , \ 
\ee
where $a$ and $b$ are positive constants, with $b<a$. It follows that
$\dot E^{(+)}_B - \dot E^{(-)}_B = -(a-b)  \sin^2 \theta$, which suggests that these vortices would be dissipated by the action of the magnetic field on time scales given by {\hbox{$T_B(\theta) \sim \rho /(\sigma B^2 \sin^2 \theta)$}}. Now, since in wall-bounded flows, vorticity lines of open vortices have their end points attached to the walls, we expect that the orientational stability of extended bulk vortices that are aligned to the magnetic field lines will depend essentially both on the wall geometric features and on the flow properties of the near-wall regions.}}

The physical picture that emerges here is that relaminarization occurs {\textcolor{black}{in the present context}} when magnetic dissipation is strong enough to hamper the growth of small-scale velocity fluctuations in the very near-wall region. As it is clear from the equality relation given in (\ref{TC}), convective and viscous dissipative processes have the same time scale $T_C$ at the top of the viscous layer. To render the argument more specific, recall the van Driest expression for the mean velocity profile in turbulent boundary layers \cite{pope, vandriest},
 \be
 u^+(y^+) = \int_0^{y^+} \frac{2 d y'}{1+\sqrt{1+4\ell_m(y')^2}} \ , \ \label{vd}
 \ee
where $y^+ \equiv y/ \ell$, $u^+ \equiv u/u_\tau$, and $\ell_m(y')$ is the Prandtl mixing length at height $y'$ 
(in viscous length units) modulated by the van Driest damping function, viz.,
\be
\ell_m(y')  = \kappa y' [1-\exp(-y'/A)] \ , \
\ee
with $\kappa = 0.41$ and $A=26$. Introduce, now, as a way to quantify the relative importance of magnetic and viscous forces, 
the height-dependent {\it{sliding Hartmann number}}, $M^\star (y^+)$, which in the units of Eq. (\ref{mhd1}) is
\bea
&& M^\star (y^+) \equiv 
\sqrt{ \frac{\sigma \left | \left ( \langle \vec v \rangle \times \vec B \right ) \times \vec B \right | }{ \mu \left | \nabla^2 \langle \vec v \rangle   \right | }} \nonumber \\
&& = \frac{M}{2 R_\tau} \sqrt{ \frac{ u^+(y^+) }{ \left. \frac{d^2}{d \xi^2} u^+(\xi) \right |_{\xi = y^+} }} 
\ . \ \label{MR_tau}
\eea
It turns out, from (\ref{vd}), that the RHS of (\ref{MR_tau}) has a single minimum at $y^+ = y^+_0 \simeq 5.5$, 
for fixed $M/R_\tau$. 

We remark that the viscous layer is usually defined as the region $y^+ < 5 \simeq y^+_0$. 
A condition for the annihilation of coherent structures produced by shear instabilities in the viscous layer -- the seeds of bulk turbulence --
can be put forward, therefore, as $M^\star (y_0^+) > C$, for some critical parameter $C$. This implies, due to the properties of (\ref{MR_tau}), 
that $M^\star (y^+) > C$ for any $y^+$, which indicates the damping action of the magnetic field over the {\textcolor{black}{viscous and buffer layers}} \cite{pope}. 
We are led, then, to the conjecture that turbulence is suppressed if
\be
\frac{M}{R_\tau} > 2 C \sqrt{ \frac{ \left. \frac{d^2}{d \xi^2} u^+(\xi) \right |_{\xi = y^+_0} }{  u^+(y^+_0)  }} \ . \ \label{MR_tauc}
\ee

\begin{figure}[ht]
\hspace{0.0cm} \includegraphics[width=0.45\textwidth]{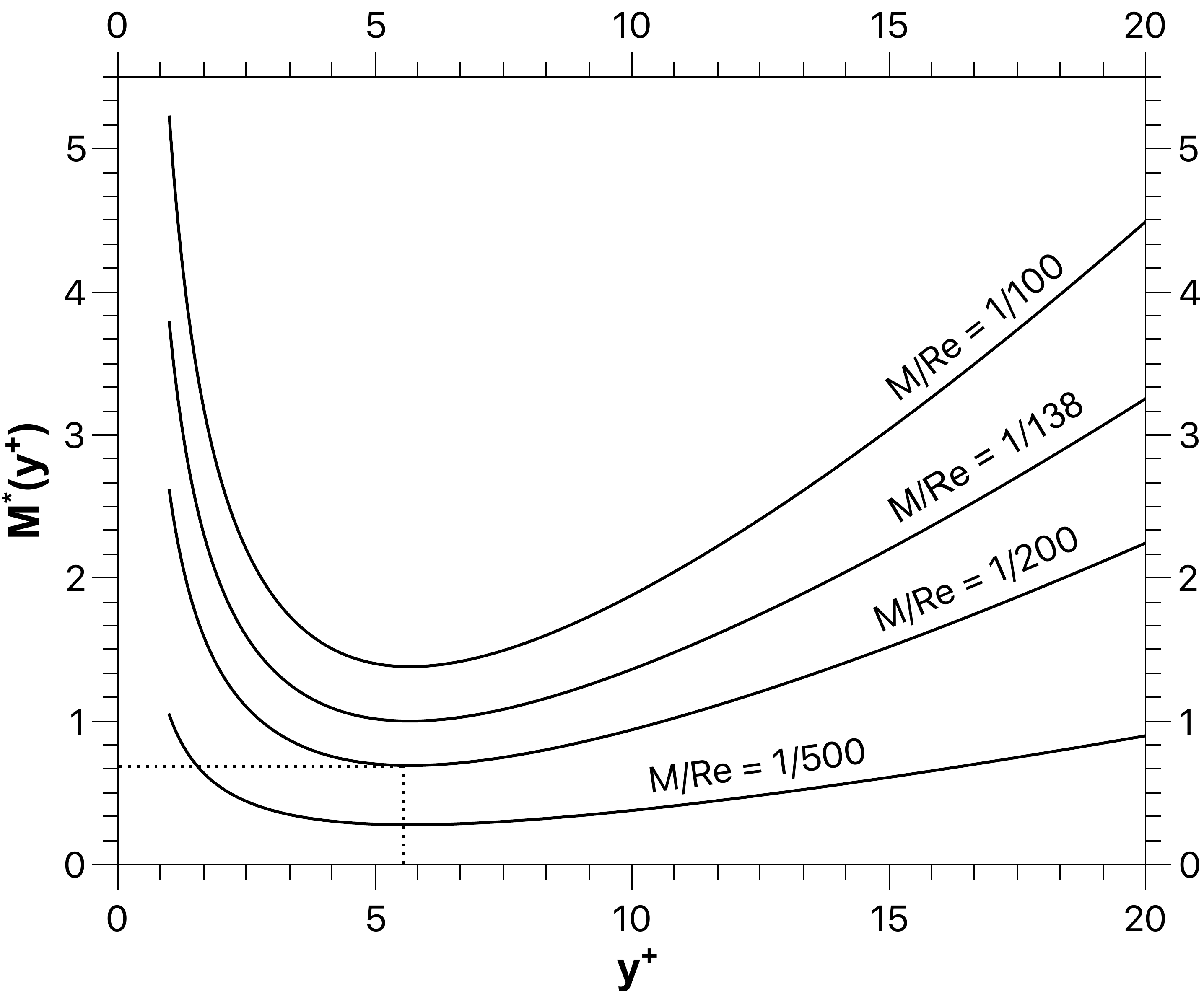}
\vspace{-0.0cm}
\caption{Sliding Hartmann numbers $M^*(y^+)$ for various ratios $M/Re$, with $Re = 10^4$. The dotted lines
give the position of the minimum of $M^*(y^+)$ for $M/Re= 1/200$.}
\label{vr}
\end{figure}

Using the standard definition of the friction factor $f$ in terms of $u_\tau$ and $U$ \cite{pope}, 
\be
f = 8 \left (\frac{u_\tau}{U} \right )^2  \ , \ \label{f}
\ee
Eq. (\ref{MR_tau}) is rewritten as
\be
M^\ast(y^+) = \frac{M}{R_e} \sqrt{ \frac{8}{f} \frac{ u^+(y^+) }{ \left. \frac{d^2}{d \xi^2} u^+(\xi) \right |_{\xi = y^+} }} \ . \ \label{MR_tau2}
\ee
As it is well-known, the friction factor depends uniquely on the Reynolds number (at fixed pipe 
relative roughness). A useful relation for the smooth pipe case is the Prandtl's empirical formula
\be
\frac{1}{\sqrt{f}} = 2 \log_{10} (Re \sqrt{f} ) - 0.8 \ , \ \label{prandtl}
\ee
which yields pragmatically accurate results for a large range of turbulent Reynolds numbers \cite{pope,mckeon_etal}. Taking $Re=10^4$ in Eq. (\ref{MR_tau2}), where $f$ is evaluated from (\ref{prandtl}), we depict, in {\hbox{Fig. 2}}, graphs of $M^\ast(y^+)$ for various values of $M/Re$.
A close look at experimental data \cite{brou_lyko, gardner_lyko} shows that relaminarization is produced, for $Re=10^4$, at {\hbox{$M/Re \simeq 1/200$}}. In this case, as it is indicated in {\hbox{Fig. 2}},  $M^\ast(y^+_0) \simeq 2/3$. According to (\ref{MR_tauc}), identifying $C$ to $M^\ast(y^+_0)$, we expect to have laminar flow for
\be
\frac{M}{R_\tau} > 0.16 \ . \ \label{critical_MRtau}
\ee
{\textcolor{black}{At this point, it is important to stress that the inner structure of turbulent boundary layers subject to transverse magnetic fields has been noticed to be reasonably well described by standard hydrodynamic turbulent boundary layer phenomenology \cite{alboussiere_lingwood}. This gives support to the heuristic arguments based on Eqs. (\ref{vd}) and (\ref{prandtl}) in determining the critical criterion (\ref{critical_MRtau}).}}

\begin{figure}[t]
\hspace{0.0cm} {\bf{(a)}} \includegraphics[width=0.45\textwidth]{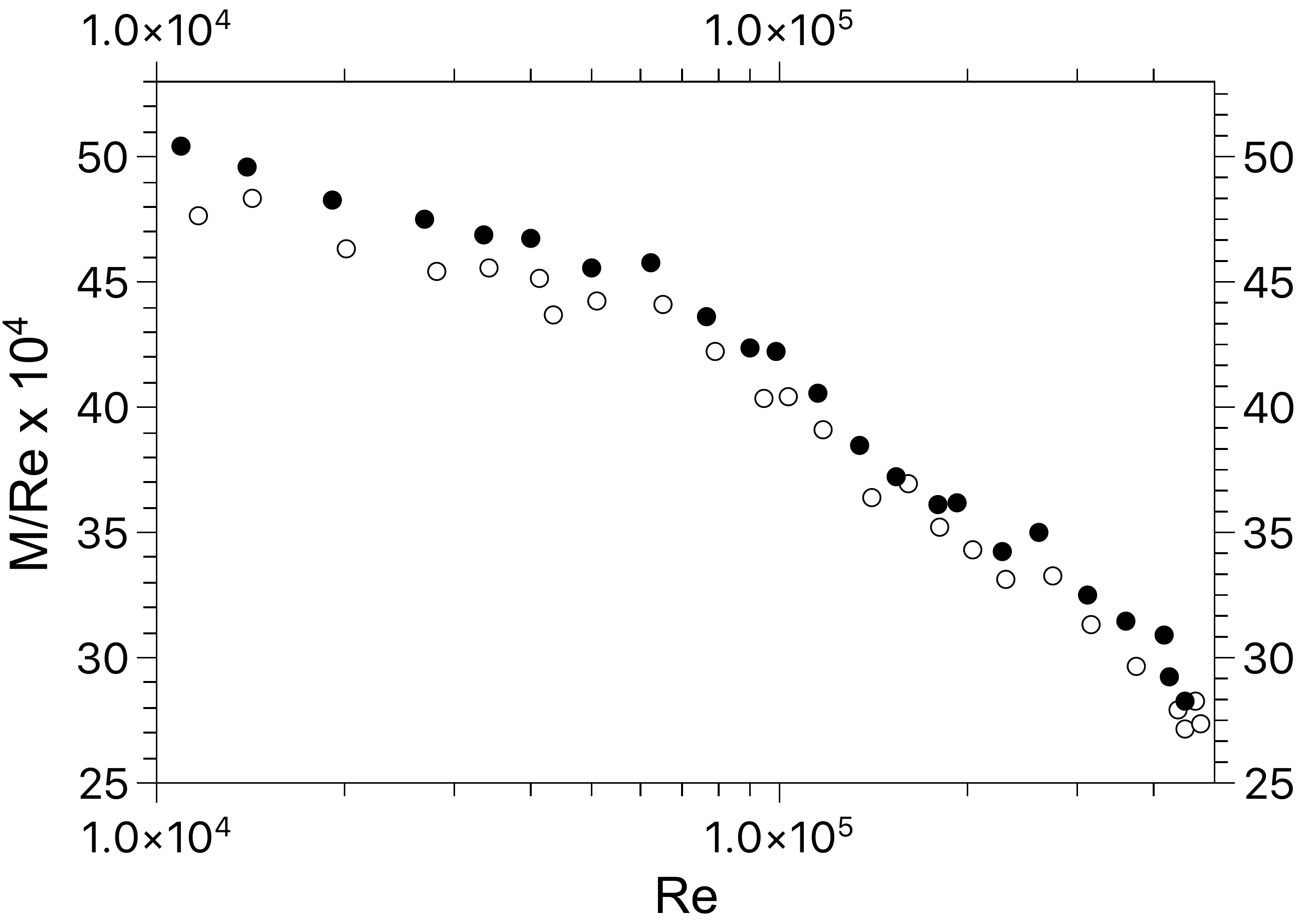}
\vspace{0.2cm}

\hspace{0.0cm} {\bf{(b)}}  \includegraphics[width=0.45\textwidth]{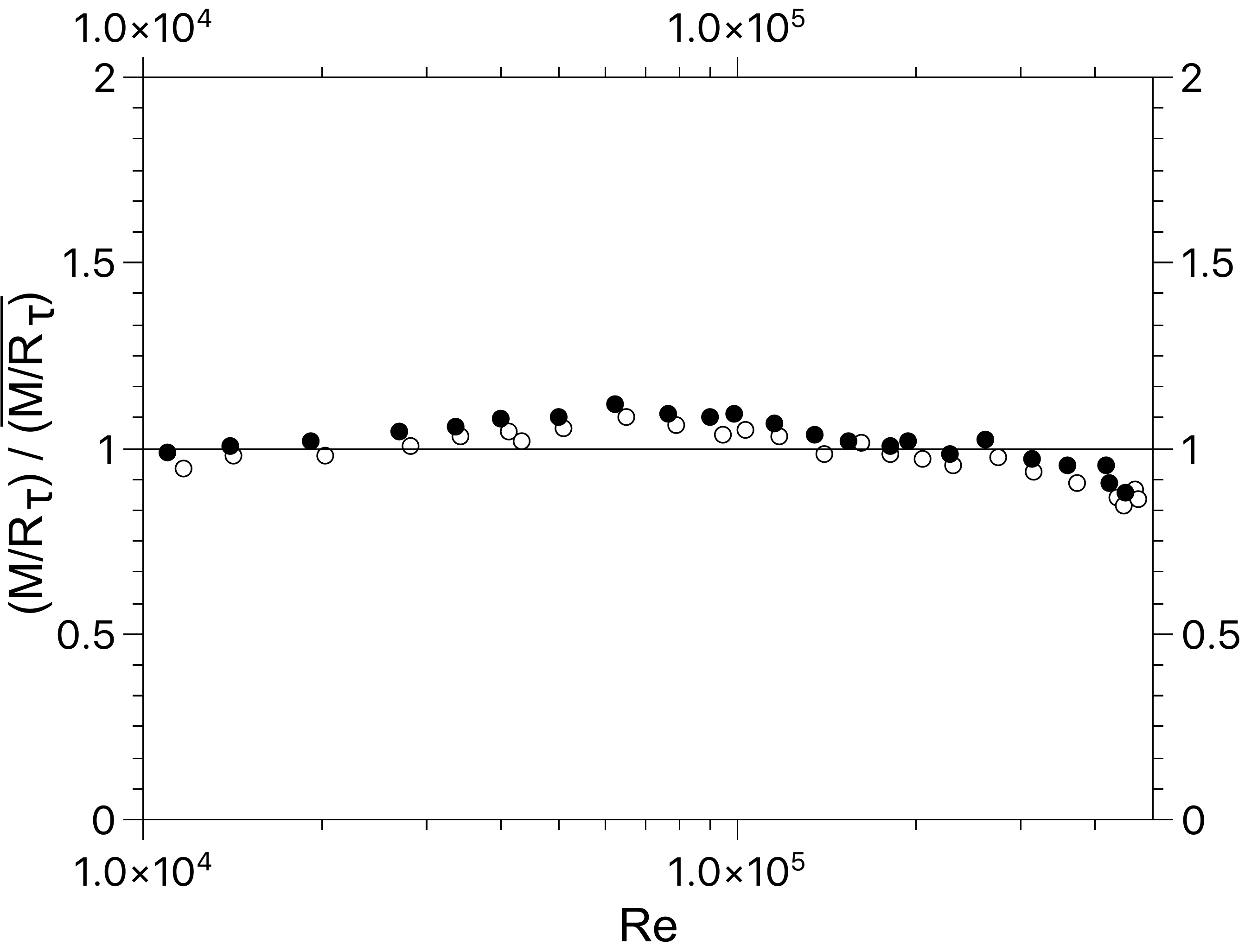}
\caption{(a) Open and closed circles refer, respectively, to the values of $M/Re$ obtained by Gardner and 
Lykoudis \cite{gardner_lyko}, for the laminar-to-turbulent and turbulent-to-laminar transitions. 
(b) Values of $M/R_\tau$, normalized by the mean value of the measured sample, as a function of the Reynolds number $Re$. These values are obtained from the Hartmann numbers extracted from the data given in (a), and from the friction Reynolds numbers evaluated through Eqs. (\ref{reynolds}), (\ref{Rt}), (\ref{f}), and (\ref{prandtl}).}
\label{vr}
\end{figure}


To check the constancy of $M/R_\tau$ at the laminar-turbulent transition for various Reynolds numbers, we explore the critical values of $M/Re$ recorded by Gardner and Lykoudis \cite{gardner_lyko}. Their values were obtained through visual inspection of the streamwise velocity signals produced by hot-film anemometry at many different radial and axial angular positions of the pipe's cross section, up to the minimum distance to the wall $\Delta^+ = 0.03 R_\tau$, so that, approximately, {\textcolor{black}{$10 < \Delta^+ < 285$}}, for the covered range of Reynolds numbers. The Gardner-Lykoudis data, in monolog scale, is shown in {\hbox{Fig. 3a}}. {\textcolor{black}{As discussed in Ref. \cite{gardner_lyko}, measurements were very sensitive to a number of conditions, like the casual oxidation and orientation of the hot-films with respect to the flow direction, as well as the temperature of the fluid (mercury). The reported data is, thus, probably not free from systematic errors, which we just hope are not too large to compromise their relevance in face of modeling attempts.}}

The measured critical values of $M/Re$ change by approximately $20 \%$ around their mean, as the Reynolds number is varied from $10^4$ to {\textcolor{black}{$4.7 \times 10^5$}}. Working instead with critical values of $M/R_\tau$, we find that their variations {\textcolor{black}{drop to below $6.5 \%$}} around the mean, as it can be inspected from {\hbox{Fig. 3b}}. 

\textcolor{black}{These results support the idea that the magnetic suppression of near-wall coherent structures is a relevant ingredient, to great extent dominant, in the process of pipe flow relaminarization for the range of investigated Reynolds numbers. It is likely, however, that further mechanisms are necessary to describe more accurately the turbulent-laminar transition as observed in the Gardner-Lykoudis pipe flow experiments. Noticing, as reported in {\hbox{Fig. 3b}}, that values of $M/R_\tau$ seem to progressively deviate from the putative constant critical line at higher Reynolds numbers (after a change of slope around $Re \simeq 6 \times 10^4$), one could conjecture that additional instabilities are triggered as the Reynolds number grows, making the point for an alternative phenomenological picture of the relaminarization process.} 

It is interesting to note that $M/R_\tau$ yields, actually, a suggestive parameterization of the effects of external magnetic fields on wall-bounded flows: $M/R_\tau$ is just the ratio of the viscous length $\ell$ of the ``would-be turbulent boundary layer" at high Reynolds numbers and negligible magnetic fields to the thickness of the ``would-be laminar Hartmann boundary layer", $R/M$, at high magnetic fields and small Reynolds numbers.

Our analysis has been restricted to the problem of relaminarization where the inlet 
turbulent flow has been previously produced without magnetic forcing, a boundary condition of 
practical relevance in many applications. It is not obvious at all, however, if at 
asymptotic far distances from the inlet the same critical relation (\ref{critical_MRtau}) between the Hartmann 
and Reynolds numbers would hold for a prediction of the laminar-turbulent transition. Loop 
experiments \cite{thess_albou}, which are likely to be related to this issue, seem to indicate that this is not so. In other words, while (\ref{critical_MRtau}) would still imply in laminar flow, laminar {\it{asymptotic}} regimes could be induced by the action of magnetic fields of relatively lower intensity, for the reported range of Reynolds numbers, $Re < 5 \times 10^5$.

To conclude, we emphasize that the phenomenological arguments carried along the above lines \textcolor{black}{-- which are essentially based on aspects of purely hydrodynamic turbulence --} are likely to
apply to the specific case of turbulent pipe flows that evolve toward regions that contain uniform transverse magnetic fields. 
It is worth pointing out, as a hint for future validation studies, that channel flows with Reynolds numbers close to the largest ones investigated in the Gardner-Lykoudis experiments are within the present reach of direct numerical simulations \cite{yamamoto_tsuji}. It is, furthermore, an interesting open issue to understand the limitations of the physical picture of laminarization induced by the magnetic suppression of near wall structures in a number of well reported alternative (non-laminar) flow regimes observed at high Hartmann numbers \cite{eckert,potherat-klein2017,potherat-kornet,suko_etal,zika_etal}.


\section{Acknowledgments}
The author thanks A. Barreto, D. Cruz, A. Freire, F. Ramos, E. Soares, and members of the PRIMATE 
(Pipe Rig for the Investigation of Magnetically Affected Turbulence in Electrolytes) Collaboration,  
D.J.C. Dennis, R. J\"ackel, J. Loureiro, and B. Magacho, for several interesting discussions. This work has been partially supported by CNPq and Petrobras (COPPETEC project number 20459).



\begin{references}

\bibitem{moreau} R. Moreau, {\it{Magnetohydrodynamics}}, Kluwer Academic Press (1990).

\bibitem{knaepen} B. Knaepen and R. Moreau, Annu. Rev. Fluid Mech. {\bf{40}}, 40 (2008).

\bibitem{davidson} P.A. Davidson {\it{Introduction to Magnetohydrodynamics}}, Cambridge University Press (2017).

\bibitem{verma} M.K. Verma, Rep. Prog. Phys. {\bf{80}}, 087001 (2017).

\bibitem{hart_laza} J. Hartmann and F. Lazarus, Mat. Fys. Medd. K. Dan. Vidensk. Selsk. {\bf{15}}, 1 (1937).

\bibitem{num1} D. Lee and H. Choi, J. Fluid Mech. {\bf{439}}, 367 (2001);

\bibitem{num2} H. Kobayashi, Phys. Fluids {\bf{20}}, 01502 (2008); 

\bibitem{num3} R. Chaudhary, S.P. Vanka, and B.G. Thomas, Phys. Fluids {\bf{22}}, 075102 (2010);

\bibitem{num4} D. Krasnov, O. Zikanov, and T. Boeck, J. Fluid Mech. {\bf{704}}, 421 (2012); 

\bibitem{num5} O. Zikanov, D. Krasnov, T. Boeck, A. Tess, and M. Rossi, App. Mech. Rev. {\bf{66}}, 03082-1 (2014).

\bibitem{tsinober1} A.B. Tsinober, {\it{MHD Flow Drag Reduction}} in {\it{Viscous Drag Reduction in Boundary Layers}},
edited by J.N. Hefner and D.M. Bushnell, American Institute of Aeronautics and Astronautics, Inc. Washington DC (1990).

\bibitem{spacecraft} L. Kai, L. Jun, and L. Weiqiang, Acta Ast. {\bf{136}}, 248 (2017).


\bibitem{scrystal} K. Hoshikawa, Jpn. J. Appl. Phys. {\bf{21}}, L545 (1982).

\bibitem{chen_etal} Q. Chen, Y. Jiang, and J. Yan, M. Qin Prog. Nat. Sci. {\bf{18}}, 1465 (2008).

\bibitem{tritiumb} H. Moriyama, A. Sagara, S. Tanaka, R.W. Moir, and D.K.Sze,
Fusion Eng. Des. {\bf{39–40}} 627 (1998).

\bibitem{ihli_etal} T. Ihli, T.K. Basu, L.M. Giancarli, S. Konishi, S. Malang, F. Najmabadi,
S. Nishio, A.R. Raffray, C.V.S. Rao, A. Sagara, and Y. Wu, Fusion Eng. Des.  {\bf{83}}, 912 (2008).

\bibitem{davidson2} P.A. Davidson, Annu. Rev. Fluid Mech. {\bf{31}}, 273 (1999).

\bibitem{steelc} R. Chaudhary, B.G. Thomas, and S.P. Vanka, Metallurg. Mater. Trans. B {\bf{43}}, 532 (2012).

\bibitem{eckert} S. Eckert, G. Gerbeth, W. Witke, and H. Langenbrunner, Int. J. Heat Fluid Fl. {\bf{22}}, 358 (2001).

\bibitem{potherat-klein2017} A. Pothérat and R. Klein,
Phys. Rev. Fluids {\bf{2}}, 063702 (2017).

\bibitem{potherat-kornet} A. Pothérat and K. Kornet, J. Fluid Mech. {\bf{783}}, 605 (2015).

\bibitem{murga} W. Murgatroyd, Phil. Mag. {\bf{44}}, 1348 (1953).

\bibitem{lykoudis} P.S. Lykoudis, Rev. Mod. Phys. {\bf{32}}, 796 (1960).

\bibitem{brou_lyko} E.C. Brouillette and P.S. Lykoudis, Phys. Fluids {\bf{10}}, 995 (1967).

\bibitem{gardner_lyko} R.A. Gardner and P.S. Lykoudis, J. Fluid Mech. {\bf{47}}, 737 (1971).

\bibitem{tsi} A.B. Tsinober, Magn. Gidrodin. {\bf{11}}, 7 (1975).


\bibitem{bran} H. Branover, {\it{Magnetohydrodynamic Flow in Ducts}}, John Wiley $\&$ Sons, New York (1978).

\bibitem{nara} R. Narasimha {\it{Relaminarization—Magnetohydrodynamic and Otherwise}} in {\it{Liquid-Metal Flows
and Magnetohydrodynamics}}, Edited by H. Branover, P.S. Lykoudis, and A. Yakhot, American Institute of Aeronautics and Astronautics, Inc. New York (1983).

\bibitem{townsend1} A.A. Townsend, Math. Proc. Camb. Philos. Soc. {\bf{47}},  375 (1951).

\bibitem{theo} T. Theodorsen,  {\it{Mechanism of Turbulence}} in {\it{Second International Midwest Conference
on Fluid Mechanics}}, Ohio State University, Columbus (1952).

\bibitem{townsend} A.A. Townsend, {\it{The structure of turbulent shear flow}}, Cambridge
University Press (1976).

\bibitem{perry_chong} A.E. Perry and M. Chong, J. Fluid Mech. {\bf{119}}, 173 (1982).

\bibitem{perry_marusic} A.E. Perry and I. Marusic, J. Fluid Mech. {\bf{298}}, 361 (1995).

\bibitem{marusic_perry} I. Marusic and A.E. Perry, J. Fluid Mech. {\bf{298}}, 398 (1995).

\bibitem{review1} D.J.C. Dennis, Ann. Braz. Acad. Sci. {\bf{87}}, 1161 (2015).

\bibitem{review2} J. Jim\'enez, J. Fluid Mech. {\bf{842}}, P1 (2018).

\bibitem{review3} I. Marusic and J.P. Monty, Annu. Rev. Fluid. Mech. {\bf{51}}, 49 (2019).








\bibitem{tobias_etal} S.M. Tobias and F. Cattaneo,
Phys. Rev. Lett. {\bf{101}}, 125003 (2008).

\vspace{10cm}


\bibitem{liu_etal} P. Liu, S.P. Vanka, and B.G. Thomas, J. Fluid Eng. {\bf{136}}, 121201-1 (2014).

\bibitem{loureiro_etal} N.F. Loureiro and S. Boldyrev,
Phys. Rev. Lett. {\bf{118}}, 245101 (2017).

\bibitem{camporeale_etal} E. Camporeale, L. Sorriso-Valvo, F. Califano, and A. Retinò
Phys. Rev. Lett. {\bf{120}}, 125101 (2018).

\bibitem{sommeria-moreau1982} J. Sommeria and R. Moreau, J. Fluid Mech. {\bf{118}}, 607 (1982).

\bibitem{davidson1997} P.A. Davidson, J. Fluid Mech. {\bf{336}}, 123 (1997).

\bibitem{potherat-klein2014} A. Pothérat and R. Klein, J. Fluid Mech. {\bf{761}}, 168 (2014).

\bibitem{alboussiere_etal1996} T. Alboussière,
J.P. Garandet, and R. Moreau, Phys. Fluids {\bf{8}}, 2215 (1996).

\bibitem{moreau_etal2010} R. Moreau, S. Smolentsev, and S. Cuevas, PMC Phys. B {\bf{3}:3} (2010).

\bibitem{suko_etal} S. Sukoriansky, I. Zilberman, and H. Branover,
Exp. Fluids {\bf{4}}, 11 (1986).

\bibitem{zika_etal} O. Zikanov, D. Krasnov, T. Boeck,
and S. Sukoriansky, J. Fluid Mech. {\bf{867}}, 661 (2019).

\bibitem{adrian2007} R.J. Adrian, Phys. Fluids {\bf{19}}, 041301-1 (2007).

\bibitem{jimenez} J. Jim\'enez, Phys. Fluids {\bf{25}}, 101302 (2013).

\bibitem{elsas_moriconi} J.H. Elsas and L. Moriconi, Phys. Fluids {\bf{29}}, 015101 (2017).

\bibitem{geng_etal} C. Geng, G. He, Y. Wang, C. Xu, A. Lozano-Dur\'an, and J.M. Wallace,
Phys. Fluids {\bf{27}}, 025111 (2015).

\bibitem{pope} S.B. Pope, {\it{Turbulent Flows}}, Cambridge University Press (2000).

\bibitem{vandriest} E.R. Van Driest, J. Aeronaut. Sci. {\bf{23}}, 1007 (1956).

\bibitem{mckeon_etal} B.J. McKeon, M.V. Zagarola, and A.J. Smits, J. Fluid Mech. {\bf{538}}, 429 (2005).

\bibitem{alboussiere_lingwood} T. Alboussière and R.J. Lingwood, Phys. Fluid {\bf{12}}, 1535 (2000).

\bibitem{thess_albou} P. Moresco and T. Alboussi\`ere, J. Fluid Mech. {\bf{504}}, 167 (2004).

\bibitem{yamamoto_tsuji} Y. Yamamoto and Y. Tsuji, Phys. Rev. Fluids {\bf{3}}, 012602(R) (2018).

\end{references}
\end{document}